\documentclass[runningheads]{llncs}

\usepackage{graphicx}
\usepackage{wrapfig}
\usepackage{longtable}

\begin{document}
\title{A Classification \\Framework for Stablecoin Designs}
%
%

\author{Amani Moin \and
Emin G{\"u}n Sirer \and
Kevin Sekniqi}
%
%

\institute{Cornell University and AVA Labs} 
\maketitle              
\begin{abstract}
Stablecoins promise to bridge fiat currencies with the world of cryptocurrencies. They provide a way for users to take advantage of the benefits of digital currencies, such as ability to transfer assets over the internet, provide assurance on minting schedules and scarcity, and enable new asset classes, while also partially mitigating their volatility risks. In this paper, we systematically discuss general design, decompose existing stablecoins into various component design elements, explore their strengths and drawbacks, and identify future directions.

\keywords{Stablecoins - Stable Payments}
\end{abstract}
\section{Introduction}
Cryptoasset prices are famous for their volatility. Though many cryptoassets aspire to become world currencies, most are frequently dismissed as no more than speculative assets due to their wild price swings. 

Money is supposed to have three functions: a store of value, a unit of account, and a medium of exchange. 
Stability is key to all these functions.
Store of value is the most salient; if people store their wealth in an asset that constantly fluctuates in value, their wealth will fluctuate accordingly. 
A volatile asset is also a poor unit of account, because it is inconvenient to denominate prices in something which constantly changes in value. 
Every time the value of the unit of account changes, all prices must be adjusted accordingly. 
Finally, and most crucially, a currency needs to be stable to function as a medium of exchange; this allows people to be be fairly and predictably compensated for goods and services without changes in value during the payment process. 

Stablecoins are a class of cryptoassets created to address this problem. 
As the name implies, they are designed to be price stable with respect to some reference point. 
There has recently been an explosion in the number of stablecoin projects announced, especially following the crash in Bitcoin prices in early 2018. 
There are over a hundred stablecoins in existence or in progress, with the top three projects now representing a market capitalization of $\$4.6$B~\cite{cmcs}.
Although the sheer number of projects seems overwhelming, they can all be decomposed into a few key features. 

\paragraph{Roadmap.}
We briefly review related works in Section 2.  In Section 3, we break down the taxonomy of stablecoins based on the first three constituent axes: the peg type, the collateral type, and the collateral amount. In Section 4, we expand on additional axes by discussing the mechanism of action chosen by various stablecoin families. In Section 5 we discuss methods to measure prices. In Section 6, we discuss some design features relevant to digital currencies in general, but especially important for stablecoins. Finally, we discuss future directions for stablecoins in Section 7.

\section{Related Work}
One of the first stablecoin taxonomies classified stablecoin projects by collateral type and discussed pros and cons of each category~\cite{qureshi}.  
Several papers and reports have followed a similar taxonomy, adding more detail on individual projects~\cite{ref1}~\cite{ref2}~\cite{ref3}.  
A paper by Pernice et. al takes a different approach, categorizing stablecoins by monetary and exchange rate regimes~\cite{ref4}. 
Our contribution is extending the existing taxonomies with a discussion of other important stablecoin design aspects, namely price stabilizing mechanisms and price measurement methods.  
We also categorize many of the existing stablecoin projects according to our extended taxonomy.

\begin{figure}[h!]
  \begin{center}
    \includegraphics[width=0.7\textwidth]{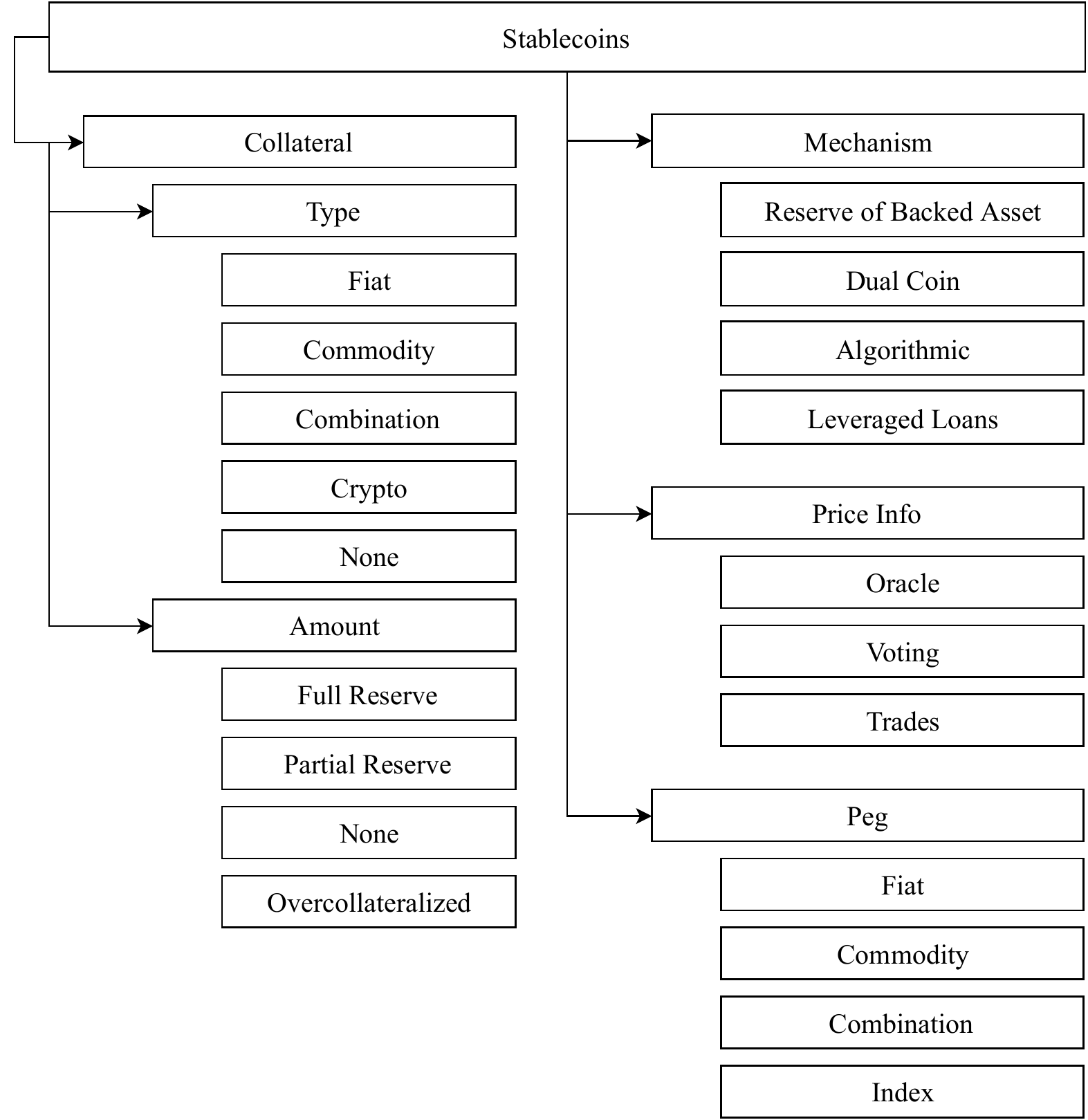}
  \end{center}
  \caption{Stablecoin taxonomy, decomposed into four main axes: peg, collateral, mechanism, and price information.}
  \label{fig:taxonomy}
\end{figure}

\section{Peg and Collateral}
\subsection{Peg}
The most salient choice for stablecoin design is the peg, which oftentimes is included in the name of the stablecoin.\footnote{Examples include TrueUSD~\cite{trueusd}, USDC~\cite{usdc}, USDX~\cite{usdx}, USDVault~\cite{usdvault}, A-Eurs~\cite{eurs}, and many others.}
USD is a popular choice, likely due to USD being typically considered a stable store of value around the world. In fact, it is not uncommon for foreign citizens, especially those in emerging and developing economies, to store their wealth in USD rather than their national currency. 
The other benefit of using USD is that price comparison is easy. 
A singular fiat currency peg allows one to check whether the peg holds by simply comparing the dollar price of an object to the pegged coin price of the same object.  
Other stable fiat currencies, such as the Euro, the Japanese Yen, and the Swiss Franc, are also popular choices for similar reasons.

Besides fiat, there are also stablecoins pegged to commodities, most commonly gold.
Some examples include Digix~\cite{digix} and HelloGold~\cite{hellogold}. 
It is interesting to note that, in general, there are fewer commodity-pegged coins than fiat-pegged coins. 
A possible explanation is that commodity prices fluctuate in value more than fiat currencies, although typically less severely than most digital currencies.

Other stablecoins may choose to peg to a bundle of currencies and/or commodities. 
This has the benefit of insulating the stablecoin against shocks to any one country,  currency, or commodity. However, pegging to a bundle can also have the opposite effect and introduce noise if some of the assets included in the bundle are very volatile.
Saga~\cite{saga}, for example, is pegged to the IMF's special drawing rights (SDR), a basket of world currencies curated by the IMF. Currencies are selected into the SDR if the issuing country is one of the world's top exporters, the currency is widely used in international transactions, and the currency is widely traded in foreign exchange markets.  However, the SDR is seldom used in any context other than the IMF's store of value and unit of account, making it a less practical choice than the dollar. Facebook's upcoming Libra also plans to peg its currency to an as of yet undetermined basket of currencies and assets. 

Saga plans to later peg their currency to the consumer price index (CPI) if they outgrow the SDR, i.e. if they become a dominant world currency. The CPI is a unitless index which tracks the inflation of the price of a basket of consumer goods. No stablecoin is currently pegged to the CPI, so it is unclear how this would be executed. It is possible, for example, that the stablecoin supply would be adjusted so the nominal price level remains constant.  Pegging to a fiat currency or commodity with finite supply can eventually lead to problems of scale, and pegging to an index can circumvent this problem.\footnote{This is one of the reasons the US went off of the gold standard.}  However, the choice of CPI as a peg is not ideal for a variety of reasons. It is typically measured monthly or even less frequently, due to logistical challenges in determining what should be in the basket and how much each component should be weighted. There are also regional differences in consumption, so it is unclear how to construct a basket that reflects global spending patterns.

\subsection{Collateral}
Emergent currencies often make use of collateral to ensure that the circulating currency has redemption value. 
This provides a lower bound on the price, thereby mitigating some of the risk of holding, using, and denominating debts in the currency. 
Since the goal of collateralizing is to bound the redemption value, it is easiest and most effective, but not necessary,\footnote{USDVault for example is pegged to USD but collateralized with gold.} to use whatever the stablecoin is pegged to. 
If users can always redeem one unit of the stablecoin for one dollar, arbitrageurs should ensure that it never trades at any other price. 

Unfortunately, collateralizing a coin creates the problem of securely storing large quantities of the collateral.
Traditionally, the best place to store large quantities of cash is in a bank, because it is secure, relatively easy to audit, and often comes with deposit insurance. 
However, this is also centralized, thus making it prone to deceptive practices. For example, Tether~\cite{tether} recently admitted it was only 74\% collateralized~\cite{tethaff}, despite initially claiming full collateralization~\cite{tether}.  Moreover, there are often limits to how much deposit insurance covers, potentially leaving the majority of the reserve uninsured.
Some stablecoins avoid this problem by storing their collateral as physical cash in a vault instead of a bank. 
For example, Rockz~\cite{Rockz} stores 90\% of its collateral in the form of physical fiat currency in an underground vault in the Swiss Alps.  

Commodity backed stablecoins also suffer from the problem of where to store their collateral, since there are fewer institutions which accept and insure deposits in the form of commodities than ones that accept cash.
This, in turn, leads to a high degree of centralization.

One way to avoid having to store large amounts of fiat is to collateralize with another cryptocurrency. 
This has the advantage of potentially decentralized operation, and allows for easier diversification across backing assets. 
The problem with this approach is that digital collateral can itself be very volatile, making it hard to use as a guarantee of value. 
Any stablecoin backed by cryptocurrencies must have some mechanism built in to safely handle large swings in the value of the underlying collateral. We discuss these mechanisms in section 4.

\begin{figure}
  \begin{center}
    \includegraphics[width=0.9\textwidth]{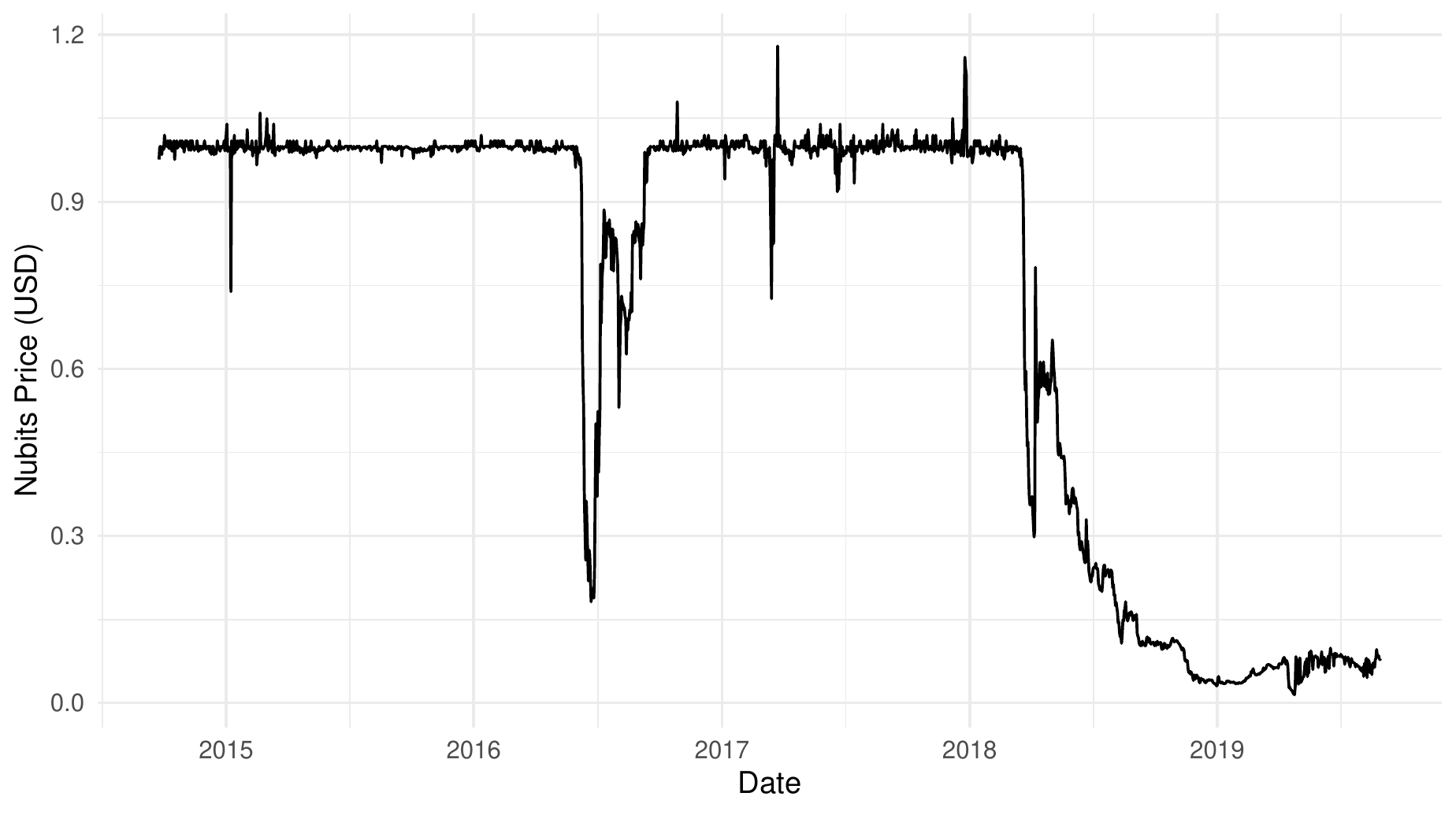}
  \end{center}
  \caption{NuBits price collapse}
    \label{fig:nubits }
\end{figure}

Other stablecoins do away with the problem of volatile collateral by simply not collateralizing the currency at all. 
This has many advantages. First, not having any collateral to store or unlock simplifies many logistical challenges. 
Second, it is also cheap to operate, since it does not require the issuer to keep real or crypto assets on hand. 
Unfortunately, this ease of operation comes with drawbacks.
Algorithms are usually gameable. 
The value of the currency in this case stems purely from the reliability of the issuing mechanism and/or people's beliefs.  Once users' expectations of the coin's stability change, whether due to a flaw in design or idiosyncratic changes in sentiment, there may be little to keep the price afloat because there is no inherent redemption value. Consequently, when these stablecoins fail, they tend to do so swiftly and catastrophically.  One example is NuBits~\cite{nubits}, which dropped from its pegged price of \$1 to less than \$0.30 over the course of 2 weeks in early 2018.  It never recovered its peg, and has been trading below \$0.10 for the past six months.\footnote{Note that USD is not collateralized, and yet it remains stable. 
However, when the dollar was a fledgling currency it was backed by gold.
It was only after extensive global adoption that the backing was gradually eased-off. 
Additionally, the US government has the infrastructure to support this type of regime. 
US federal law makes it so that businesses are required to accept US issued currency as legal tender. 
There are also regulatory and executive agencies that enforce compliance.}
 
\subsection{Collateral amount}

Hand in hand with the decision of collateral type comes the decision of collateral amount. 
Since collateral serves to support the price by creating a reliable redemption value, the best choice seems to be a fully collateralized stablecoin. 
If every unit of currency can be redeemed for the underlying asset, there is virtually no rational reason they should ever trade at different prices, thus making price fluctuations minimal.  
However, one to one collateralization is sometimes excessive, and sometimes insufficient. 

Having a full reserve, where the value of the collateral is exactly the value of the circulating currency, makes it hard for a currency to scale. 
As the stablecoin becomes more widely used, the issuers have to keep buying more collateral in order to keep up with demand. 
Nevertheless, this stablecoin design has been successfully utilized by Hong Kong's currency board; the Hong Kong Dollar is fully collateralized by USD and has maintained a roughly 7.8 to 1 peg to the US dollar since the early 1980s. It is currently the 13th most traded currency in the world~\cite{hkd}.    

Instead of staying fully collateralized, some currencies, like Saga, try to mimic the historical trajectory taken by the US Dollar. 
Such currencies initially fully collateralize their stablecoin, then slowly reduce their collateral ratio and ease off the peg once the money supply has exceeded some threshold. 
Although Tether eventually admitted they were not fully collateralized as they initially claimed, there was no ostensible detriment to the price. 
A reason for the continued stability despite only partial collateralization is that full collateral is not necessary as long as people do not believe that more than the entire reserve amount will ever be cashed out at once. 
It is also worth noting that almost any supposedly fully collateralized fiat backed stablecoin whose collateral is being held in a bank, such as USDC, is functionally a partial reserve currency. All commercial banks keep only some of their deposits on hand and use the rest for investments or to issue loans.  This does not cause any problems as long as they keep enough on hand to satisfy demand for withdrawals.

Other coins, especially algorithmic ones such as Basis~\cite{basis}, do not keep any collateral at all. 
Instead, value is preserved purely by expanding supply when the price is too high and contracting it when the price is too low.  
On the other end of the spectrum, many currencies collateralized by crypto-currencies keep more than the value of the circulating currency in reserve to guard against price swings in the collateral. 
This way, even if the collateral asset depreciates, there is still enough for each unit of the stablecoin to be redeemed for an equivalent amount of the underlying asset. 
 
\section{Mechanics}
All stablecoins require some mechanism to adjust the price when it deviates from the peg. 
Usually, this is done by expanding supply when the price is too high and contracting it when the price is too low. 
This means that there usually needs to be some way of measuring the price (covered further in the next section) and knowing how much to expand or contract the supply. 
Most stablecoins are designed such that rational, self interested users will act to restore the peg when the price deviates. For example, this could be achieved by allowing users to redeem stablecoins for collateral when the price of the stablecoin is too low. 
Other stablecoins issue a secondary token designed to absorb the volatility of the first, resulting in a stablecoin/volatilecoin pair. 
Still others depend on an algorithmic market making mechanism or central-bank contract to manage the supply. 
 
\subsubsection*{Reserve of Pegged Asset} 
Many stablecoins will build a mechanism where users will be incentivized to expand or contract the supply until the price returns to the peg. 
The simplest way to achieve this is in a fully collateralized system backed by the pegged asset, and allow users to expand supply when the price is too high and redeem when the price is too low. 
Arbitrageurs earn money while helping maintain the peg. 
For example, if a stablecoin pegged to USD is trading at less than \$1, stablecoin holders should redeem the coin for the underlying collateral, thereby buying a dollar for less than a dollar.
This will contract the supply until the price returns to the peg and the arbitrage opportunity disappears, or until the reserve runs out. 

On the other hand, if the market price of the stablecoin is above \$1, many systems will allow users to expand the supply by wiring funds to the account where the rest of the collateral is being held. 
This allows the user to buy something worth more than \$1 by paying only \$1 for it. 
The simplicity and autonomy of this system makes it extremely appealing, which is why a majority of stablecoins in circulation today use this method, or a very similar one.  However, it is not foolproof.  On October 15, 2018, the price of Tether briefly dropped below \$0.93 due to a large selloff.  The price recovered to above \$0.98 within the day and appears to have suffered no lasting effects~\cite{cmcs}.  Other notable examples of this design include USDC, TrueUSD, Carbon~\cite{carbonmoney}, Paxos~\cite{pax}, Gemini Dollars~\cite{geminidollar}, and many others.

As stated previously, the main problem with allowing users to always redeem for collateral is storing large amounts of collateral at some physical location. 
Since it is expensive to provide the security needed to protect large sums of money, most stablecoins rely on one central location, like a bank. 
This introduces two issues: 
centralization and dependency on legacy financial institutions. 
USDC gets around one of these problems by storing collateral at a network of banks rather than a single one. 
However, it still relies heavily on existing banks. 
The other problem with this type of system is the ability to scale. 
As previously discussed, this makes it difficult, though not impossible, to become a global currency due to the inconvenience of storing assets of such high value in a large number of locations. This inconvenience, is one of the reasons USD outgrew the gold standard.

A common variation on this design requires a central authority to mint the coins, but allows people to redeem the stablecoin for the underlying collateral. 
This creates a lower bound on the price of the stablecoin but not an upper bound, since users can redeem when the stablecoin price is too low but cannot mint when the price is too high. 
This is common in cases where the collateral is not necessarily dollars, such as Digix. 
Since it would be inconvenient to accept and verify gold deposits from individual users, users are not allowed to mint Digix by contributing capital to the collateral pool. 
They can, however, still redeem their Digix for physical gold, thereby creating a lower bound on the value of Digix. 

In addition to allowing people to mint coins, Tether was released in waves, allegedly at strategic times to prop up the price of Bitcoin~\cite{tethprop}. 
Users can redeem Tether for USD, but since Tether is no longer fully collateralized, they also reserve the right to deny people the right to redeem.

Another variation being employed by Facebook's Libra~\cite{libra} is to allow only the set of validators to mint or redeem coins, instead of all users.  This reduces overhead since presumably larger amounts would be transacted each time, and at a lower frequency.  This may come at the cost of a lower speed of adjustment, since the set of potential arbitrageurs who can correct the price is restricted.

\subsubsection*{Dual coin} Another way to maintain stability is to pair the pegged coin with a secondary coin which absorbs the volatility of the first. 
The most well known example of this is the seigniorage shares model employed by the original formulation of Carbon. 
When the price of the stablecoin dips below the peg, a secondary coin is auctioned in exchange for the stablecoin. 
The proceeds from the auction are then burned to contract the supply. 
When the price of the stablecoin is above the peg, additional coins are minted to holders of the secondary token. 
Holders of the secondary token help prop up the price when the currency inflates and are rewarded during deflationary periods. 

There are two big concerns with this type of system. 
One is that the secondary coin often meets the SEC's definition of a security. 
Regulatory complications stemming from this designation were enough to keep Basis from launching~\cite{basisclose}. 
Carbon also changed from a dual coin system to a fiat backed system, for undisclosed reasons, possibly due to regulatory hurdles. 
The second concern is that if holders of the primary token do not believe that the stablecoin will appreciate in the future, there is no incentive to buy or hold the secondary token.  
In other words, one needs a strong contingent of users who, even during a downturn, believe that the stablecoin will eventually appreciate in value.  Additionally, since cryptocurrency markets are often subject to long downturns, people may be reluctant to wait for extended, indeterminate amounts of time for their investment to pay off.  Since there is no collateral backing this system, if people are not willing to buy the secondary coin, there will be no force propping up the value of the stablecoin.

Variations on this design include USDX and Celo~\cite{celo}, which use concepts from dual coin systems and redemption based systems to peg their tokens. 
USDX is a stablecoin collateralized with Lighthouse (LHT), a digital currency. 
USDX is pegged to USD but backed over 200\% by LHT, which is stored in two centralized funds. 
People are free to exchange LHT to USDX and vice versa at a valuation of \$1/USDX. 
As such, when USDX is trading at a price of less than \$1, people are incentivized to redeem it for \$1 worth of LHT, contracting the supply and raising the price. 
LHT ends up absorbing the volatility of USDX because the supply of USDX is contracted by expanding the circulating supply of LHT. 
The main issue in this design stems from USDX being 200\% collateralized by LHT. The largest possible market cap for USDX is half the market cap of LHT, and if the market value of LHT declines, so does the potential market cap of USDX. It is unclear what LHT derives its value from, so the potential market cap of USDX might be small and/or unstable. 

Celo works similarly to USDX, but with a few additions. 
Celo is backed by CeloGold,\footnote{Despite its name, CeloGold not actually backed by gold or any other asset} Bitcoin, and Ethereum. 
In addition to allowing people to redeem Celo for CeloGold and vice versa, there is an algorithmic central bank which buys and sells Celo and CeloGold to stabilize the price.  
 
StatiCoin/RiskCoin~\cite{staticoin} is another variation of the dual coin system which allows creation and redemption of coins to maintain a peg. 
Users who want to mint StatiCoin send ETH to the contract and receive a dollar equivalent amount of StatiCoin in return. 
When they want to redeem, they send StatiCoin to the contract and receive an equivalent dollar amount of ETH. 
Users who want to mint RiskCoin also send ETH to the contract and receive a corresponding amount of RiskCoin in return, depending on what the current RiskCoin price is. 
StatiCoin can always be redeemed for \$1 worth of ETH, provided there is enough collateral, while RiskCoin holders are the residual claimant to the ETH held by the contract. 
If there is \$100 worth of ETH held by the contract, 90 StatiCoins in circulation, and 2 RiskCoins in circulation, the value of all of the outstanding Riskcoin is \$10 and each RiskCoin is worth \$5. 
If the next day the value of the ETH held in the contract drops to \$90, RiskCoin will be worth \$0 and users will not be allowed to mint additional RiskCoin until the dollar value of collateral exceeds the number of StatiCoin issued. 
StatiCoin is unique because it is crypto-collateralized but does not require overcollateralization, an inefficient mechanism for absorbing volatility. 
If the value of the collateral falls below the market cap of StatiCoin, StatiCoin may become unpegged because not all holders of StatiCoin will be able to redeem for the underlying collateral.  If the price of ETH drops enough to reduce the price of RiskCoin to 0, then StatiCoin will become an unreliable store of value precisely when Ethereum is losing value and a stable valued asset is most needed. 
 
Terra \cite{terra} is yet another dual coin stablecoin, with its volatility absorbed by the secondary token Luna. 
Luna serves two purposes in this system.  
It absorbs volatility because it is auctioned to contract the supply when the stablecoin price is too low and purchased with Terra when the stablecoin price is too high. 
It is also the staking token of the system. 
Fees and mining rewards, paid in Terra, are increased when Terra decreases in value to incentivize holding Luna and smooth the pro-cyclicality of the value of Luna. 
Unlike previously discussed designs, such as Staticoin/Riskcoin, the secondary token Luna can retain value and creates payoffs from fees and mining rewards even in an extended contraction. 
However, it does this by raising fees when the value of Terra is decreasing, which discourages Terra use when the value of Terra is already low. 
Terra also increases the staking pool in times when the market cap of Terra is small and there is less need for stakers, and decreases it when the market cap is large and more security is needed. 

Yet another variation on the dual coin system is the triple coin system proposed by Basis. Instead of having a singular volatility absorbing coin, Basis has bond tokens and share tokens. Bond tokens are auctioned off when the price of Basis decreases below \$1, and each one is redeemable for 1 Basis when the price of Basis is above \$1.  If all of the Bond tokens have been redeemed and the price of Basis is still above \$1, additional Basis is minted pro rata to holders of share tokens until the price of Basis returns to \$1.  Holders of Basis bonds absorb the downside and are the first to benefit from increases in the market cap of Basis, while holders of Basis shares benefit only when there are large expansions in the value of Basis.  There are infinite ways to tranche the volatility absorbing coin; in theory there could be systems with four or more coins splitting the stablecoin volatility across several parties.

\subsubsection{Algorithmic} Other currencies use a fully algorithmic approach to adjust the supply of the stablecoin in response to price fluctuations. 
One such example is Ampleforth, previously named Fragments \cite{ampleforth}. 
Whenever the value of Ampleforth changes, token holders have their balances adjusted proportionally to preserve the value of a single token. 
For example, if Ampleforth is originally worth \$1, then, after an increase of 10\% to \$1.10, all balances will automatically be inflated by 10\%. 
Likewise, if the value of Ampleforth declines, each Ampleforth holder's balance will be decreased accordingly. 
This makes Ampleforth a stable unit of account, since by design, the ratio of Ampleforth's market cap to the number of Ampleforth tokens is periodically adjusted to be \$1. 
Unfortunately, this is not a good store of value. 
Holding Ampleforth is no different than holding a non-pegged coin: if the market cap of Ampleforth declines, users' balances and outstanding payments will decline proportionally. 

A different algorithmic approach is employed by Saga. 
Although Saga does not peg the value of its coin, it uses an algorithmic path-independent market maker inspired by Bancor~\cite{bancor} to provide liquidity and to dampen sudden price fluctuations. 
The market maker sets the price and bid ask spread for Saga based on how much collateral it has in its reserve. 
For example, when Saga is just starting out and its reserve is small, the price will be set at 1 SDR, and the market maker will sell a Saga for 1.0015 SDR and buy for 0.9985 SDR.  This makes it so that users should not sell Saga on secondary markets for less than 0.9985 or buy for more than 1.0015 SDR, which limits how suddenly the price can change.  As the reserve grows and shrinks, the price and spread are gradually adjusted in response. 
Like Ampleforth, Saga does not guarantee that the value of Saga holdings will be stable over time.  However, the market maker does guard against sudden price movements, and thus provides short-term stability.

\subsubsection*{Leveraged loans} Leveraged loans are a system of stablecoins which utilize components from all the above classes.  Dai \cite{dai} is the most successful example of such a system. 
Users lock up collateral, such as Ethereum and other cryptoassets, in collateralized debt positions (CDPs). 
They can then mint Dai, a stablecoin pegged to \$1, up to 2/3 the value of the collateral in the CDP. 
Users can then unlock their collateral by paying back the borrowed Dai, plus a stability fee that accrues over time. 
Dai is destroyed once it is paid back. 

If the value of the collateral in a CDP drops below 1.5x the Dai borrowed, the debt position is automatically liquidated, and the collateral is used to purchase the amount of Dai borrowed against it. Any remaining collateral, minus a liquidation fee, is returned to the original CDP owner. 
If the value of the collateral depreciates quickly and drops below the value of the Dai borrowed, a secondary coin is minted to cover the difference. Currently, the secondary coin is PETH, but eventually it will be transitioned to MKR.
Since MKR is the governance token, and MKR holders are diluted when CDPs are underwater, there is an incentive for the holders of the governance token to set parameters such that users are not defaulting on their loans. However, we note that the probability of the value of the collateral declining to less than the Dai borrowed is low since the price of the collateral would have to suddenly drop by over 33\%.

Users are incentivized to buy Dai and unlock their collateral when the price of Dai decreases, because a decrease in the price of Dai makes it cheaper for them to unlock their collateral. This contracts the supply and restores the peg.  If Dai continues to trade at a price lower than its intended peg, MKR holders can vote to raise the stability fee charged to CDP holders. This serves as further incentive for CDP holders to liquidate their positions and contract supply. 

Dai received a lot of attention in March 2019 for consistently trading around \$0.98 instead of \$1 as it was supposed to.  Since then, it underwent a series of stability fee increases, some of which quixotically lowered the price of Dai instead of raising it as intended. 
Despite this issue and Ethereum's price decrease by  $\sim90\%$ since Dai launched, Dai has managed to remain within $\sim2\%$ of its pegged value.

 \begin{figure}[h!]
    \centering
    \includegraphics[height=0.3\textheight]{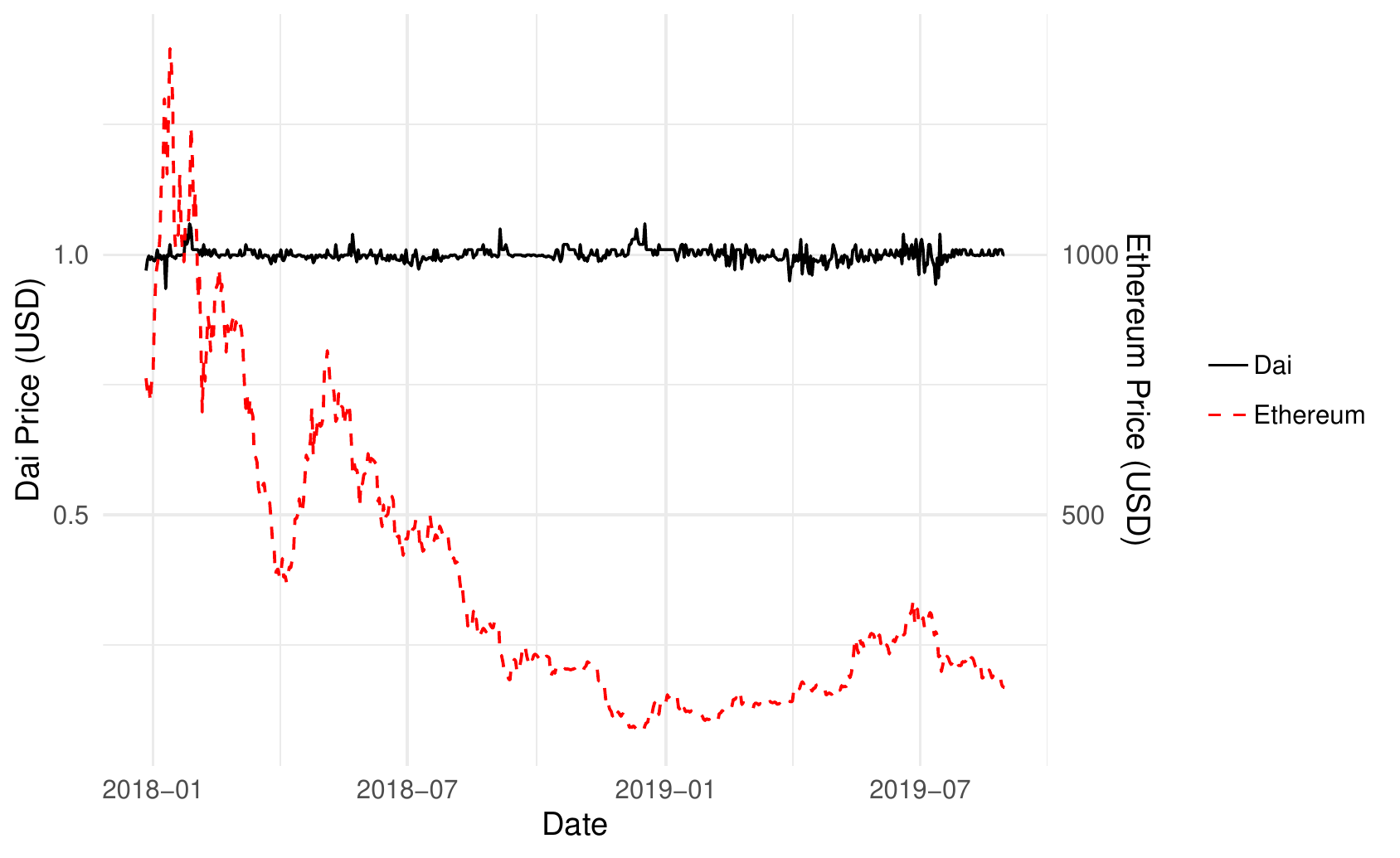}
    \caption{DAI remains relatively price stable despite decline in ETH price.}
    \label{fig:daip}
\end{figure} 
 
\subsubsection*{Miscellaneous}
There are a few other designs that do not neatly fit into any of the above categories. For example, Steem~\cite{steemdollars} props up the price of its stablecoin, Steem Dollars, by paying interest on Steem Dollars. However, since they don't set negative interest rates, this mechanism may not work if the price of Steem is too high and the interest rate is already low.  Consequently, Steem Dollars traded for more than its \$1 peg for months despite an interest rate of 0\%.  

Another design is employed by NuBits (now defunct), a stablecoin which is minted when holders of a secondary coin (NuShares) vote to create more.  Users are also paid interest if they temporarily remove their NuBits from circulation.  This project is no longer in operation, possibly due to voting on supply changes being a slow process, thus forcing adjustments in price to lag by several days or more.  Additionally, if holders of NuShares also hold NuBits, they may be reluctant to dilute the value of NuBits by printing more. 

Kowala~\cite{kowala} keeps the price stable by adjusting its mining rewards. When the price of the stablecoin is too high, rewards increase to dilute the supply; when the price of the stablecoin is too low, transaction fees are burned to contract the supply.  Unfortunately, a decline in the price of the stablecoin might be correlated with fewer transactions occurring, since a break from the peg would diminish users' confidence in the stablecoin. Since price adjustments are effected through mining and transactions, recovering from a decrease in price would take a long time. Also, since mining rewards decrease during contractionary periods, miners have less incentive to provide security which may further diminish the value of Kowala.  This could lead to a feedback loop where Kowala never recovers from a price decrease. 

Finally, Phi~\cite{phi} offers people the opportunity to issue loans denominated in Phi, a stablecoin.  The loan issuer has to put up collateral, which is used to pay the loan if the borrower defaults.  Although the issuer does collect interest on the loan, the issuer has have no way to recover their capital if the borrower defaults. Moreover, because there is no connection to real world identities, there is no ostensible consequence to defaulting, so borrowers will likely abscond with the loan.  If the borrower does pay back the loan, they are supposed to pay it with interest denominated in Phi. Since, for every loan originated in Phi, the amount paid exceeds the amount created, there could be not enough coins in existence to pay back all existing loans with interest. Finally, although Phi serves as a unit of account for the loans issued, it is not clear why the value should remain stable, or why Phi might be used as a store of value or medium of exchange. 

\section{Price information}
A crucial step in making supply adjustments at the appropriate times is accurately measuring the price. 
Most stablecoins make use of an external oracle, an independent price feed(s) deemed trustworthy by the issuers of the stablecoin. 
This leaves a crucial component of the system completely out of the hands of the stablecoin issuer. 
The entities publishing the price feed might deviate from their standard practice in how they calculate prices and trigger disastrous downturns or upturns for the stablecoin. 
This is not unheard of, since, for example, CoinMarketCap suddenly and abruptly decided to stop including prices from exchanges in South Korea, resulting in a sudden drop in reported prices~\cite{cmckorea}.
If a stablecoin was formerly using this price feed and desired no change in how prices are calculated, the system would be left with few options other than to accept the new price feed, find a different oracle, or adjust oracle prices to correct for the new calculation method.  Short term pricing errors can arise from using an oracle too, as was the case with Synthetix. In June 2019, a commercial API used by Synthetix suffered a glitch and began to report incorrect exchange rates, resulting in a bot making over \$1B during this period~\cite{synthetix}.  Although the bot owner chose to reverse the trades during this episode, there is no guarantee that the next profiteer will be as generous.  Note that these are examples which arose even with no malicious adversaries in the system.  

On the other hand, if there is a malicious actor intent on sabotaging the stability, a price oracle can serve as a potential target.  Increasing the number of price feeds might be a potential solution to this issue. However using the median makes price updating slow, since the system must wait for sufficient majority of price feed reports. As a result, even some of the most active and popular stablecoins, including MakerDao, use only a few price oracles, making them a potential source of attack vectors~\cite{mkrtools}. 
There are only 14 price oracles for MakerDao, so hijacking of any 8 would corrupt the median-price rule. Moreover, these oracles may not be fully independent, as they might have overlap in where they obtain their price information or in their deployment platform. 

Nonetheless, the use of external oracles persists because the alternatives are generally worse.
Prices for most assets are generated based on the prices at which the assets are transacted on exchanges.
However, many crypto exchanges, both centralized and decentralized, have stale prices and/or inflation of trade volumes~\cite{bamsec}.
If trades are being inflated by the exchange, it is possible that exchanges might just be taking prices from some external feed and adding noise.
Unless the initial exchanges are chosen wisely and with near perfect foresight, a non-noisy price feed is just as good or better.

Alternately, Schelling point mechanisms~\cite{vitalikschell}, a.k.a. crowd oracles, can also be used to set the price. 
The justification for this method is that it is hard for voters to coordinate on a deceptive answer. 
However, with a pegged coin, there exists a natural alternate coordination point: the pegged price. 
If this is a more advantageous equilibrium for voters, then information obtained in this manner is not going to be trustworthy. 
Many of these schemes use rewards for being close to the median and slashing for voters far from the median to incentivize truth telling. 
However, this may incentivize people to answer how they think others will answer, commonly known in economics as the beauty pageant problem. 
Take for example Basis, a variant of the dual coin example discussed earlier, whose original design mentioned the possibility of using a crowd oracle. 
If users correctly express that the stablecoin has appreciated and is trading above its peg, more of the stablecoin will be minted, and users who only hold the stablecoin and not the secondary coin will be diluted. 
This makes it such that the payoff for holders of the stablecoin is higher if they lie and claim that the stablecoin is trading at its intended price rather than its true price. 
Even if a user wants to tell the truth, when enough people are incentivized to divert, the rest of the honest users will have to lie, abstain from voting, or be penalized.

Terra tries to get around the problem of dishonest voting by sampling only a subset of voters to make collusion difficult. 
However, if there is a non truthful equilibrium that is beneficial for a majority of voters, then the subsampling may not help. 
Celo also uses a crowd oracle and acknowledges that there is potential for price manipulation. 
The designers of Celo trust that holders of the voting token will prioritize long term growth over short term profit, which may be an incorrect assumption.

Some stablecoins are designed so that no external oracle is needed for the stablecoin to remain stable. 
In systems where users can always trade in for the underlying collateral, such as Tether or Saga, there is no need for a price feed. 
Instead, prices are measured using users' trades. 
Individual users decide how to value the token and then cash in or out accordingly. 
However, as previously discussed, the convenience of not having to measure the price usually comes at the cost of having to store collateral. 
 
\section{Other Considerations}
 
\subsubsection*{Governance}
Flexible governance is the ability to change system parameters and operation dynamically in response to changes in the environment, allowing a coin to scale or overcome unforeseen obstacles. 
This is an emerging choice for digital currencies, first popularized by Tezos.  It has since been adopted by others, including several stablecoins. 

Although the idea of crowd-sourcing system parameters caters to a democratic ideal consistent with crypto's ethos of decentralization, in actuality, participation may be low. 
This may leave important decisions in the hands of a motivated minority. 
For example, the March 7, 2019 stability fee increase in Dai was approved with less than 1\% of MKR holders voting and a single address contributing more than 50\% of the stake~\cite{mkrvote}. 
Besides low turnout, voting may suffer from high latency. 
If specific governance changes require human participation, then the process becomes highly contentious and complicated.
On the other hand, if there is no way to amend the governance, the founding team must foresee every problem or potentially hard fork every time a change has to be made. 
This can lead to systems which are inflexible and react poorly to changing global environments.

\vspace{-2ex}
\subsubsection*{Fees}
Fees can be used to incentivize good behavior, such as how Dai's liquidation fee penalizes low levels of collateral. 
The presence of this fee rewards MKR holders who are supposed to police the CDPs and penalizes CDP owners who are negligent with their balances. 
However, they can also introduce pricing frictions. 
For example, if a fully collateralized fiat backed token has no fees, arbitrageurs should guarantee that it never trades at anything but \$1. 
However, fees (such as those employed by TrueUSD) introduce a friction which prevents arbitrageurs from taking advantage of and correcting price discrepancy. 
Suppose that a coin which is supposed to trade for \$1 is instead trading for \$0.99. 
If there is a cashing out fee of \$70 (as with TrueUSD), someone would have to buy \$6930 worth of TrueUSD and cash it in at \$7000 just to break even. 
To make a profit, they would have to invest even more money into this strategy. 
This is can be capital intensive for the arbitrageur. 
Moreover, the cashing out process is not instantaneous since wire transfers take time to process. This introduces an opportunity cost as this strategy can tie up capital for a day or more each time it is used. 

\vspace{-2ex}
\subsubsection*{Regulatory Compliance}
Another aspect some coins are grappling with is the degree of regulatory compliance. 
Although KYC/AML compliance avoids the possibility of regulatory problems down the road, it also alienates some potential users.  Some people may demand absolute privacy, be concerned about secure consumer data storage, or be unbanked because they lack the paperwork necessary to go through the KYC process. Other regulatory costs, such as the time, effort, and lawyers required to file with the relevant regulatory agencies, make it prohibitively expensive to launch a coin, as was the case with Basis~\cite{basisclose}.  And finally, being KYC/AML compliant in one country does not protect the issuer from liability if the coin is being used in another country. 

It is especially crucial for coins that are fiat backed with collateral stored in banks to not break any laws because banks can freeze accounts if they suspect suspicious or illegal activity.  
For example, if the stablecoin is being abused for money laundering or other similar purposes, depository banks can stop money from being withdrawn or deposited in the account. 
This would prevent users from creating or redeeming tokens and hinder quantity adjustments necessary to keep the price stable. 
This is not a purely hypothetical problem. 
In April 2017, Tether found themselves unable to accept international wire transfers into their Singapore bank accounts and were denied outgoing wire transfers by Wells Fargo~\cite{wf}. 

\section{Future Directions}
\subsection{Stable Pay}
One alternative to stabilizing an entire currency supply is to only stabilize payments. This can be a cheaper and easier than stabilizing the entire monetary supply. The only currency currently incorporating this strategy is Xank~\cite{xank}. Payments on the Xank network have the option of being stabilized by the algorithmic central bank; the bank subtracts Xank tokens when the price of Xank increases and adds them when it declines so the dollar value stays constant.  The central bank continues to make these adjustments until the tokens are used in another transaction or the user cashes out of their Xank position.  Although such a currency can serve as a medium of exchange and, to some extent, a store of value, it cannot serve as a unit of account because the value of individual tokens is allowed to fluctuate.  The largest problem with this design is that the central bank is essentially providing a free put option, and providing a valuable service for free is a difficult business model to sustain.  Crypto markets tend to have fairly long run ups and declines, so there is a high degree of serial correlation in returns.  If users expect prices to increase, they should unpeg their transactions. If users expect prices to decrease, they will keep their payments stabilized.  In a prolonged downturn, this can lead to the central bank running out of money and being unable to continue to stabilize payments.  

\subsection{Peg to Other Assets}
Another area for stablecoin expansion is assets pegged to financial assets other than currencies, such as real estate or stock or bond indices.  This would make it easier for people to diversify their holdings across digital currencies and real assets without the inconvenience of cashing out of crypto in order to do so.  The traditional finance sector is slowly becoming more interested in crypto markets.  Bitcoin futures have been listed on the Chicago Mercantile exchange since the end of 2017.  Also, NASDAQ lists several blockchain companies and is exploring uses for blockchain technology, indicating an openness to an integration between the traditional financial sector and crypto.  Although there are several regulatory hurdles in the way, it is possible that a broader range of financial assets might eventually be available on crypto markets. 

\section{Conclusion}
Stablecoins, that is, low-volatility, programmable, and auditable currencies, promise to bridge the chasm between fiat currencies and digital currencies. 
Their importance in on-ramping the trillions of assets into digital form is evident in the sheer number of stablecoins issued over the last few years. 
In this paper, we provided a systematic overview of all the different types of stablecoins developed, and divided the various proposals into constituent design elements, based on peg, collateral type and amount, stability mechanism, and price information. Although there are hundreds of projects in existence, most are variations of the same few components.  
There is still much potential for growth in this area.  Although there are many promising designs, none are without their flaws. Further innovation will be necessary before cryptocurrencies adequately fulfill the functions of money well enough to be adopted by mainstream users.   

\bibliographystyle{splncs04}
\bibliography{main}

\clearpage
\section{Appendix A: Illustration of Stablecoin Mechanisms}

\renewcommand{\thefigure}{A\arabic{figure}}
\setcounter{figure}{0} 

 \begin{figure}[h!]
    \centering
    \includegraphics[width=0.55\linewidth]{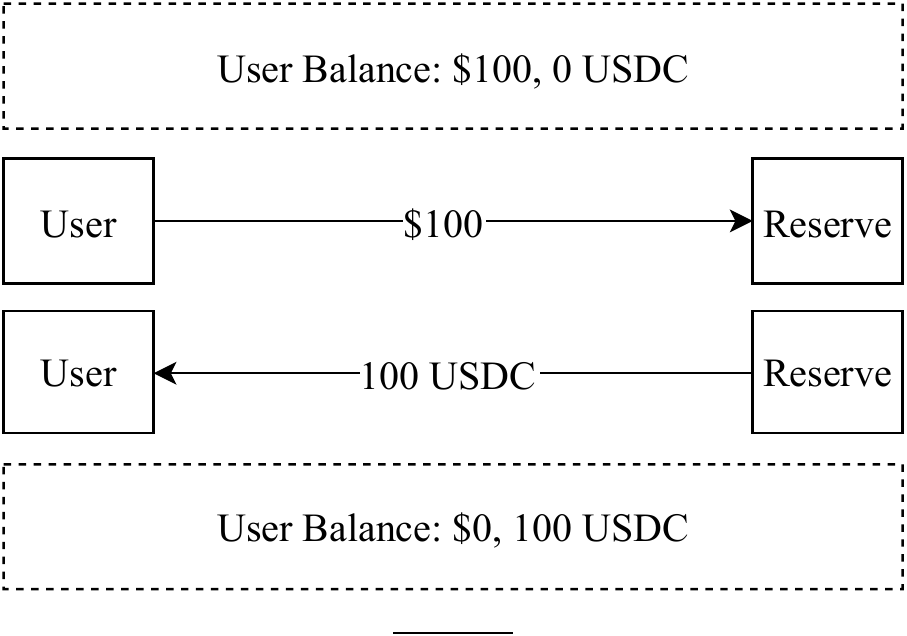}
    \caption{Reserve of Pegged Asset (ex. USDC)}
    \label{fig:tether }
\end{figure}

\begin{figure}[h!]
    \centering
    \includegraphics[width=\linewidth]{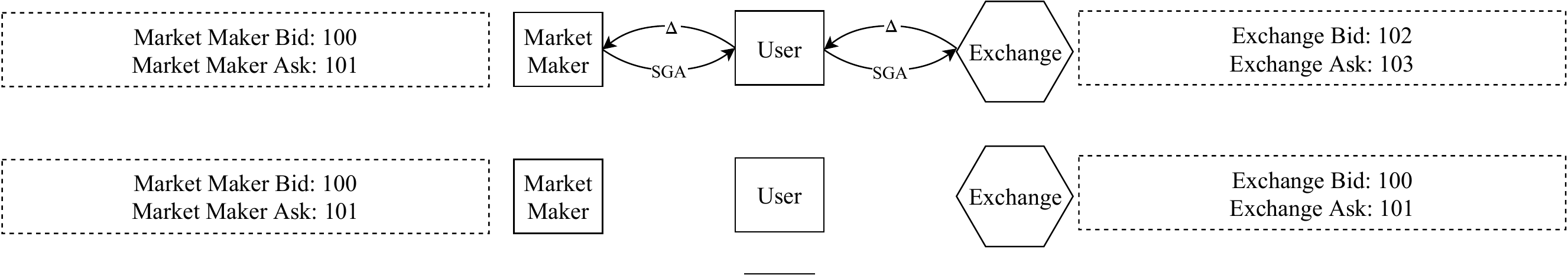}
    \caption{Algorithmic (ex. Saga)}
    \label{fig: Algorithmic}
\end{figure}

\begin{figure}[h!]
    \centering
    \begin{minipage}{0.45\textwidth}
    \includegraphics[width=0.95\textwidth]{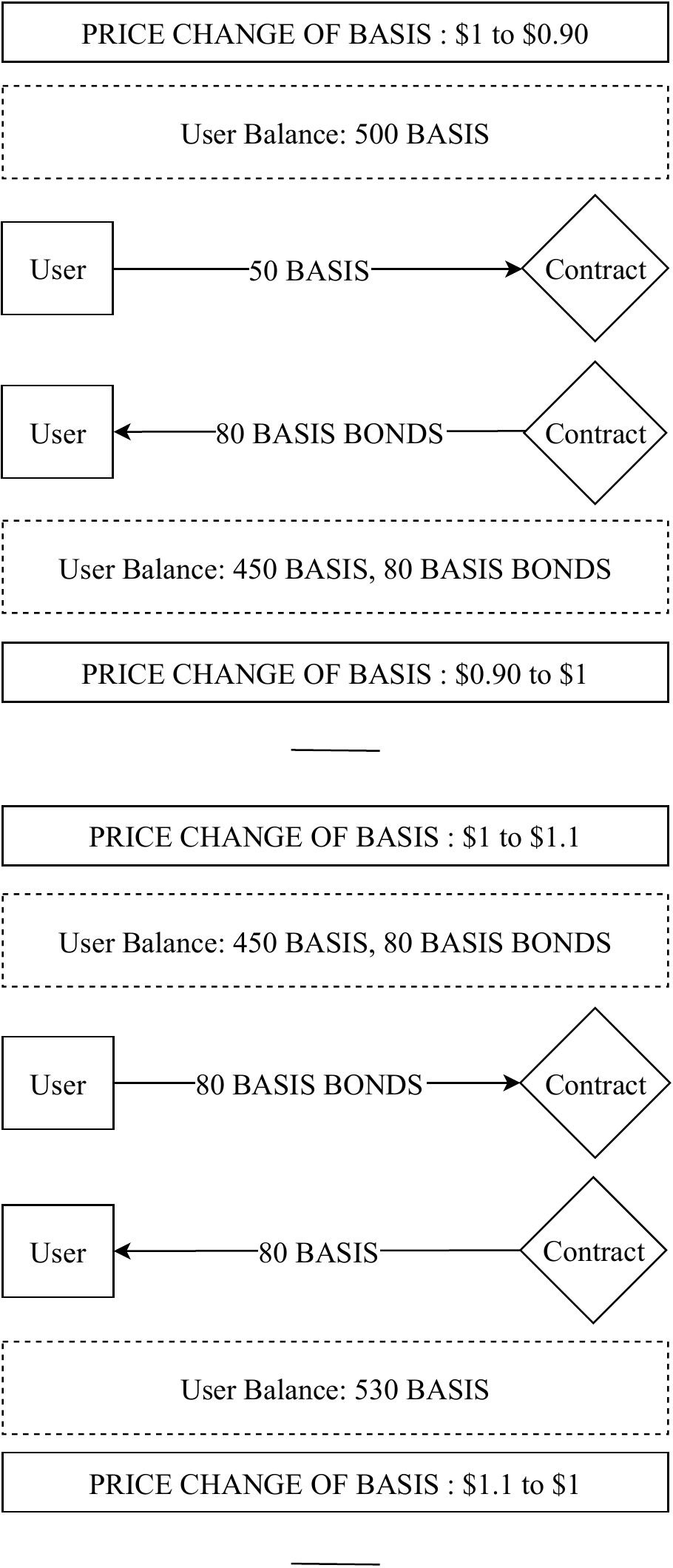}
    \caption{Dual Coin (ex. Basis)}
    \label{fig:basis}
    \end{minipage}
    \hspace{.05\linewidth}
    \begin{minipage}{0.45\textwidth}
    \includegraphics[width=0.95\textwidth]{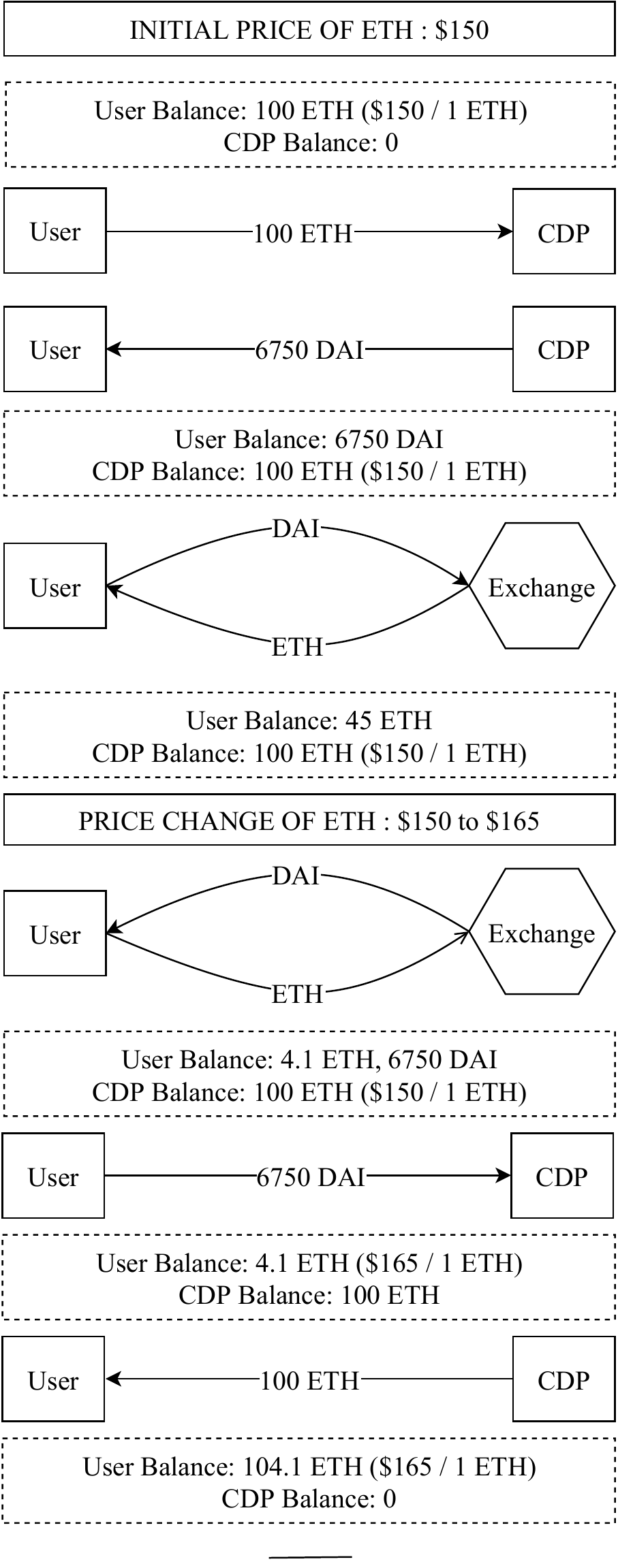}
    \caption{Leveraged Loans (ex. Dai)}
    \label{fig:dai}
    \end{minipage}
\end{figure}

\clearpage
\section*{Appendix B: Classification of Existing Stablecoins}
\setlength\LTleft{-3.2cm}
\setlength\LTright{0pt plus 1fill minus 1fill}
\begin{longtable}{| p{2.5cm} || p{1.5cm} | p{3 cm} | p{2.2 cm} | p{5.2cm} | p{3cm} |}
  \hline
  
  \textbf{Name} & \textbf{Peg} & \textbf{Collateral} & \textbf{Collateral Amount}  & \textbf{Price \& Supply Adjustments} & \textbf{Price Information}
  \\ \hline
  
  Carbon &  USD &  USD &  Full &   Reserve of pegged asset & Trades     \\ \hline

  Tether~\cite{tether}&  USD &  USD/Euro &   Partial & Reserve of pegged asset & Trades \\ \hline

  TrueUSD~\cite{trueusd} &  USD &  USD &  Full & Reserve of pegged asset & Trades \\ \hline

  Basis &  USD &  None &  None &    Two coin & Oracle \  \\ \hline

  BitUSD~\cite{bitshares} &  USD & BTS & Full or over & leveraged loans &  Elected delegates input exchange prices    \\ \hline

  Saga~\cite{saga}&  initially SDR, later CPI or none &  Basket of Fiat (SDR) & Full to partial & Algorithmic market maker & Based on amount of reserve  \\ \hline

  Bancor~\cite{bancor} &  ETH &  ETH & Over &  Algorithmic market maker &  Based on token balances  \  \\ \hline

  Dai~\cite{dai} &  USD & ETH &  Over &  Leveraged loans & Median oracle  \\ \hline

  Gemini Dollar~\cite{geminidollar} &  USD &  USD &  Full &   Reserve of pegged asset & Trades  \\ \hline

  USDC~\cite{usdc} & USD & USD & Full &  Reserve of pegged asset & Trades \\ \hline

  AAA Reserve~\cite{AAA} & Avg inflation for G-10 countries & Fiat, fixed income, and loan investments & Full & Determined by by Arc Fiduciary Ltd & Trades and oracle 
  \\ \hline

  DGX~\cite{digix} & Gold & Gold & Full & Reserve of pegged asset&Trades  \\ \hline

  EURS~\cite{eurs} & Euro & Euro & Full & Reserve of pegged asset&Trades 
  \\ \hline

  StableUSD/ Stably/ USDS~\cite{stably} & USD & USD & Full & Reserve of pegged asset&Trades   \\ \hline

  PAX~\cite{pax} & USD & USD & Full & Reserve of pegged asset&Trades   \\ \hline

  White standard~\cite{white} & USD & USD & Full & Reserve of pegged asset&Trades   \\ \hline

  SDUSD~\cite{alchemint} & USD & NEO  & Over & Leveraged loans &External oracle elected by SDS holders   \\ \hline

  JPM Coin~\cite{jpm} & USD & USD & Full & Reserve of pegged asset&Trades   \\ \hline

  USDX~\cite{usdx} & USD & Lighthouse (LHT) & 200\%+&Dual coin variant  & Median of exchange prices, validated by users 
  \\ \hline

  Stronghold USD~\cite{stronghold} & USD & USD & Full & Reserve of pegged asset&Trades  
  \\ \hline


  sUSD~\cite{nomin} & USD & Synthetix (SNX) & 5x & Leveraged loans & External oracle
  \\ \hline

  eUSD (Havven)~\cite{eusd} & USD & ETH&Over & Leveraged loans & External oracle   \\ \hline

  eUSD by Epay~\cite{eusde} & USD & USD & Full & Reserve of pegged asset & Trades 
  \\ \hline

  NuBits~\cite{nubits} & USD & None & None & Nushareholders vote whether to list more NuBits on an exchange, or offer interest to take NuBits out of circulation &Voting \\ \hline

  Token~\cite{token} & USD  & USD  & Full & Reserve of pegged asset & Trades  \\ \hline

  Monerium~\cite{monerium} & One for each major currency & Same as peg & Full & Reserve of pegged asset & Trades   \\ \hline


  Reserve~\cite{reserve} & initially USD & initially USD, later other assets & Initially full & Reserve of pegged asset & Trades 
  \\ \hline

  Terra~\cite{terra} & SDR & None&None & Dual coin & Randomly sample users who vote on price
  \\ \hline

  Ampleforth~\cite{ampleforth} & USD & None & None & Supply is expanded or contracted proportional to market cap&Whitelist of trusted oracles 
  \\ \hline

  Augmint~\cite{augmint} & Euro & ETH & 2x&Leveraged loans  & Exchanges 
  \\ \hline

  Bridgecoin~\cite{bridge} & USD & ETH  & 2x&Leveraged loans +algorithmic market maker&  Oracle 
  \\ \hline

  HelloGold~\cite{hellogold} & Gold & Gold & Full&Reserve of pegged asset & Trades 
  \\ \hline

  Kowala~\cite{kowala} & USD & None & None &Block rewards increase when price is high and are burned when price is low& Large holders of mUSD(staking token) act as oracles  \ \\ \hline

  x8c~\cite{x8c} & None & Gold, USD, Euro, GBP, JPY, AUD, CAD, NZD,CHF & Full & AI shifts funds across currencies to keep value constant&External oracle 
  \\ \hline

  NOS~\cite{nos} & USD & USD & Full & Reserve of pegged asset & Trades \\ \hline

  Phi~\cite{phi} & USD & TBD & Over & Phi is minted when validators issue a loan and burned when the loan is paid back & TBD   \\ \hline

  Celo~\cite{celo} & USD & Celo Gold, BTC, ETH & Variable & Dual coin + algorithmic central bank &Crowd oracle 
  \\ \hline

  Aurora Boreal~\cite{boreal} & USD&ETH and other reserves &Partial& Supply expands when Decentralized Capital issues loans denominated in Boreal and contracts when loans are repaid, users will also receive grants to act as market makers & Unknown\   \\ \hline

  Stableunit~\cite{stableunit} & USD & Cryptoassets & Initially over &  Algorithmic market maker&Median oracle 
  \\ \hline


  Rockz~\cite{Rockz} & CHF & CHF & Full & Reserve of pegged asset & Trades 
 \\ \hline

  Steem Dollars~\cite{steemdollars} & USD & None & None & Interest accrues to Steem Dollar holders&Steem Power holders elect oracles 
  \\ \hline

  USDVault~\cite{usdvault} & USD & Gold & Full & Reserve of pegged asset, "sophisticated gold hedging process" to maintain peg & Trades
  \\ \hline

  Globcoin.io~\cite{globcoin} & Gold and currency basket & Gold and 15 largest currencies & Full & Reserve of pegged assets, can cash out to any one currency in basket & Oracle \\ \hline

  JCash~\cite{jibrel} & USD, KRW and other assets & USD, KRW and other assets& Full & Reserve of Pegged asset&Trades\\ \hline

  Staticoin~\cite{staticoin} & USD & Eth & Full & Dual coin variant&24h exchange price 
  \\ \hline

  Unum~\cite{unum} & USD& Cryptoassets & Under to over depending on prices& Algorithmic market maker: users sell crypto or Unum to smart contract & External oracle  \\ \hline

  Poly~\cite{topl} & None & Tokenized commodities & Full & Reserve of pegged asset  &Trades 
  \\ \hline
  
 BitBay~\cite{bitbay} & None & None &None & Freeze and unfreeze tokens based on transaction and staking history. Users receive interest on frozen coins &Dynamic Peg oracle\ \\ \hline


  BitCNY~\cite{bitshares} & Chinese Yuan & BTS & Full or over & leveraged loans &  Elected delegates input exchange prices  \  \\ \hline

  EOSDT~\cite{eosdt} & USD & EOS & Over & Leveraged loans &Oracle 
  \\ \hline

  Neutral~\cite{neutral}& USD & Other stablecoins (PAX, TUSD, DAI, and USDC) & Full &Reserve of pegged asset  &Trades   \\ \hline











  Candy~\cite{candy} & Mongolian Tugrik & Mongolian Tugrik & Full &Reserve of pegged asset &Trades 
  \\ \hline



  Onegram~\cite{onegram} & Gold & Gold & Full &  Reserve of pegged asset & Trades \  \\ \hline




  Carats.io~\cite{caratsdotio} & Diamonds & Diamonds & Full &  Reserve of pegged asset & Trades \  \\ \hline




  
  Libra~\cite{libra} &Collateral basket &Bank deposits and short-term government securities &Full&Reserve of pegged asset&Trades and oracle \\ \hline
  
   Anchor~\cite{anchor} &Monetary measurement unit&None &None&Dual coin& Oracle \\ \hline

  \end{longtable}

\end{document}